# Unlocking the Potential of $Ni^{2+}$ and $Ni^{2+}$-$Cr^{3+}$ Synergy for Bifunctional Pressure and Temperature Optical Sensing

Maja Szymczak[a,*], Lukasz Marciniak[a,*]

[a] Institute of Low Temperature and Structure Research Polish Academy of Sciences, 50-422 Wroclaw, Poland

*corresponding authors: m.szymczak@intibs.pl, l.marciniak@intibs.pl

**Abstract**

Reliable simultaneous optical sensing of pressure and temperature under extreme and dynamically fluctuating conditions remains a major challenge due to intrinsic cross-sensitivity between these two thermodynamic parameters. Multimodal systems enabling simultaneous yet fully decoupled monitoring of both parameters are therefore highly sought after. Here, we demonstrate that the synergistic interplay between $Cr^{3+}$ and $Ni^{2+}$ luminescence provides a platform for bifunctional temperature-pressure sensing with independent readout channels. Two complementary detection strategies were systematically investigated: ratiometric approach based on luminescence intensity ratio and kinetic approaches exploiting emission decay dynamics. Among the kinetic strategies, a time-gated dual-ion lifetime concept - introduced here for the first time for luminescence manometry - enables pressure readout with record-high relative sensitivity reaching 148.33% $GPa^{-1}$ while exhibiting complete immunity to temperature fluctuations. Conversely, temperature sensing is achieved via time-gated single-ion $Ni^{2+}$ luminescence, ensuring high thermometric performance with negligible pressure-induced interference. Importantly, this work study, for the first time, the potential of $Ni^{2+}$ ions for application in near-infrared luminescence manometry. The unique combination of ultrahigh sensitivity, multimodal readout capability, and possibility of near-infrared operation positions the Ni$^{2+}$-Cr$^{3+}$ luminescence synergy as a benchmark platform for next-generation bifunctional optical sensors, enabling reliable operation in complex, dynamically evolving, and optically demanding environments.

## 1. Introduction

Luminescence-based optical sensing has emerged as one of the most powerful strategies for monitoring pressure and temperature in materials and devices operating under extreme and dynamically evolving conditions.[1,2] Remote, real-time readout, combined with the intrinsic chemical robustness and photophysical stability of inorganic phosphors, renders this approach particularly attractive for harsh and industrially relevant environments. Importantly, the recently demonstrated possibility of decoupling pressure and temperature readout has significantly enhanced measurement reliability, paving the way toward multifunctional optical sensing platforms.[3–8] Among pressure-responsive phosphors, Cr$^{3+}$-activated materials have become one of the most intensively investigated systems.[9–13] Their high pressure sensitivity originates from the strong dependence of the $^4T_2$ excited-state energy on crystal-field strength, which is directly modulated by pressure-induced shortening of metal-oxygen bonds. This pronounced crystal-field coupling leads to substantial pressure-driven shifts of the broad Cr$^{3+}$ emission band, enabling highly sensitive optical pressure readout. Moreover, the broadband nature of the $^4T_2 \rightarrow \, ^4A_2$ emission has enabled the development of ratiometric sensing strategies based on luminescence intensity ratios (*LIR*). In contrast, conventional pressure standards such as Al$_2$O$_3$:Cr$^{3+}$ (ruby) or SrB$_4$O$_7$:Sm$^{2+}$ rely on narrow emission lines and pressure-induced spectral shifts.[14–16] While ideally suited for high-pressure metrology in diamond anvil cells, narrowband-based sensors exhibit limited intrinsic sensitivity and are inherently unsuited for two-dimensional pressure mapping, as their readout relies on precise spectral line tracking rather than intensity- or ratio-based imaging approaches. As thermal and mechanical effects often produce comparable energy changes, reliable pressure extraction under non-isothermal conditions becomes challenging. Although recent $Cr^{3+}$-based *LIR* approaches partially mitigate this limitation, most broadband pressure sensors operate in the visible or short-wavelength near-infrared region below 800 nm.[6,10,13,17,18] The same restriction applies to $Mn^{2+}$-, $Eu^{2+}$-, and $Ce^{3+}$-based systems, which are the most commonly utilized.[19–24] In practical environments, however, phosphors are typically embedded in polymers, binders, lubricants, or coatings that exhibit strong absorption, scattering, and parasitic luminescence in the visible spectral range. These optical interferences distort spectral readout and significantly reduce sensing reliability. In this context, pressure monitoring based on luminescence kinetics remains comparatively underexplored. Lifetime-based detection relies on intrinsic excited-state dynamics and is largely independent of excitation power fluctuations, concentration variations, or optical alignment.[25,26] When combined with near-infrared (NIR) emission - where optical scattering and background luminescence[27] are significantly reduced-kinetic readout offers a powerful strategy for minimizing environmental interference. Nevertheless, NIR-operating pressure sensors exploiting luminescence kinetics remain scarce.[6,12,17] These considerations strongly motivate the development of bifunctional pressure-temperature-sensitive phosphors operating entirely in the near-infrared region. Among luminescent ions, $Ni^{2+}$ is particularly attractive in this regard, as its emission typically occurs beyond 1000 nm.[28–30] According to the Tanabe-Sugano diagram for the $d^8$ configuration, the $^3T_2$ excited state exhibits exceptional sensitivity to crystal-field variations, suggesting strong susceptibility of $Ni^{2+}$ emission to pressure-induced lattice compression.[31,32] Despite this favorable electronic structure, $Ni^{2+}$ luminescence has not yet been exploited for pressure sensing, primarily due to its intrinsically low emission intensity. Here, this limitation is overcome by employing $Cr^{3+}$ ions as sensitizers for $Ni^{2+}$ emission in co-doped $ZnGa_2O_4$, enabling efficient energy transfer and enhanced deep-NIR luminescence.[33,34] Moreover, the incorporation of $Cr^{3+}$ not only compensates for the intrinsically low emission intensity of $Ni^{2+}$, but also introduces a second optically active center within the same host lattice. The coexistence of two spectroscopically distinct dopant ions naturally enables the design of a dual-ion sensing architecture, in which each ion can serve a complementary role in the optical readout. Consequently, this strategy allows, for the first time, systematic investigation of $Ni^{2+}$-based pressure sensing while simultaneously establishing a platform for bifunctional pressure-temperature detection.

To fully exploit the synergistic interplay between $Cr^{3+}$ and $Ni^{2+}$ luminescence, two complementary pressure readout strategies were implemented: (i) a kinetics-based mode utilizing time-gated luminescence analysis, and (ii) a ratiometric *LIR*-based approach derived from the emission spectra. Both strategies were systematically investigated in two configurations: using exclusively the $Ni^{2+}$ emission and within a dual-ion architecture exploiting the $Cr^{3+}$-$Ni^{2+}$ synergy. In parallel, the same ion configuration was examined from a thermometric perspective. Crucially, this multifunctional design enables simultaneous pressure and temperature monitoring without mutual cross-sensitivity, as schematically shown in Figure 1. The dual-ion kinetics-based pressure mode remains intrinsically temperature-invariant, while the $Ni^{2+}$-based *LIR* and time-gated ratiometric mode serve as robust thermometric parameters. Notably, the time-gated dual-ion lifetime-based approach proposed here for the first time in luminescent manometry delivers a record-high relative pressure sensitivity among kinetics-based optical manometers, reaching 148.33% $GPa^{-1}$ while maintaining full temperature decoupling.

This integrated strategy enables the rational design of a bifunctional pressure-temperature sensor featuring multimodal optical readout. Importantly, the sensing platform gives opportunity to operate in the near-infrared region beyond 800 nm, minimizing optical scattering, background luminescence, and interference from encapsulation media, thereby significantly enhancing reliability under realistic operating conditions.

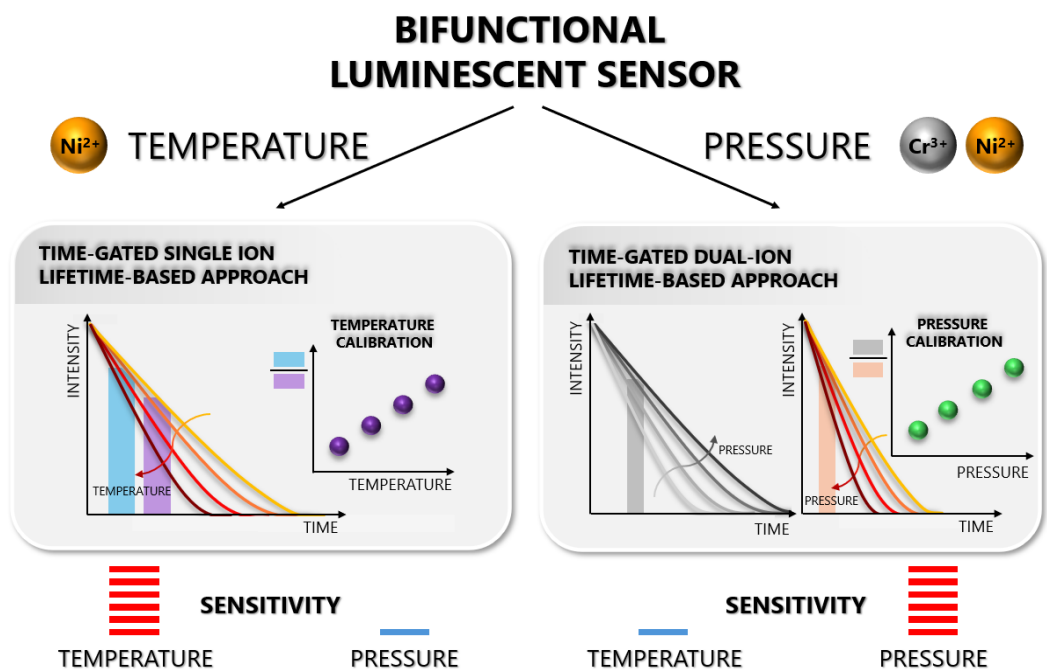

**Figure 1.** Conceptual illustration: the designed pressure sensor based on $ZnGa_2O_4:Ni^{2+},Cr^{3+}$ enables independent and selective readout of pressure and temperature by exploiting ion-specific emission channels and tailored time-gated detection strategies. Temperature sensing is achieved by monitoring the time-gated luminescence of $Ni^{2+}$ ions, whose emission dynamics exhibit high thermal sensitivity while remaining weakly affected by pressure fluctuations. Conversely, pressure readout is realized using a time-gated dual-ion lifetime approach based on the intensity ratio extracted from selected time windows of $Cr^{3+}$ and $Ni^{2+}$ emission. The combination of spectrally and temporally separated emission channels thus provides a multimodal sensing platform capable of decoupled pressure-temperature monitoring within a single material system.

## 2. Experimental Section

*Materials*

All precursors for the synthesis were used without additional purification: $Zn(NO_3)·6H_2O$ (Alfa Aesar, 99.998% of purity), $Ga_2O_3$ (Alfa Aesar, 99.999% of purity), $NiCl_2·6H_2O$ (Alfa Aesar, min. 99.995% of purity) and $Cr(NO_3)·9H_2O$ (Alfa Aesar, min. 99.99% of purity).

*Synthesis*

ZnGa$_2$O$_4$ powders doped with x%Cr$^{3+}$ ions (x = 0.1, 0.2, 0.4, 0.6, 0.8, 1) and co-doped with 0.8%Cr$^{3+}$ and x%Ni$^{2+}$ ions (x = 0.1, 0.2, 0.4, 0.6, 0.8, 1) were synthesized via a conventional high-temperature solid-state reaction route. All starting reagents were weighed in stoichiometric proportions and thoroughly homogenized in an agate mortar. A few drops of hexane were added during grinding to enhance mixing uniformity. The resulting homogeneous precursor mixtures were transferred into ceramic crucibles and calcined in air at 1573 K for 6 h with a heating rate of 10 K min$^{-1}$. After the thermal treatment, the samples were allowed to cool naturally to room temperature inside the furnace and subsequently reground to obtain fine, homogeneous powders.

*Characterization*

To evaluate the phase purity and crystal structure of the obtained samples, X-ray diffraction (XRD) measurements were performed using a PANalytical X'Pert Pro diffractometer equipped with an Anton Paar TCU1000 N temperature control unit and Ni-filtered Cu Kα radiation (40 kV, 30 mA).

The morphology of the synthesized powders and the elemental distribution within individual microparticles were examined by scanning electron microscopy (SEM) using an FEI Nova NanoSEM 230 instrument equipped with an energy-dispersive X-ray spectrometer (EDS, EDAX Genesis XM4) for elemental mapping. Prior to SEM analysis, the powders were dispersed in a small amount of methanol, and a drop of the resulting suspension was deposited onto a carbon-coated aluminum stub. The samples were then dried under an infrared lamp.

The excitation and emission spectra, as well as luminescence decay profiles, were recorded using an FLS1000 fluorescence spectrometer (Edinburgh Instruments) equipped with a 450 W xenon arc lamp (used exclusively for excitation spectra measurements) and an R5509-72 photomultiplier tube (Hamamatsu) operated in a nitrogen-flow cooled housing. For

temperature- and pressure-dependent emission spectra and decay measurements, a 445 nm laser diode was employed as the excitation source.

During temperature-dependent measurements, the sample temperature was controlled using a THMS 600 heating-cooling stage (Linkam) with a temperature stability of 0.1 K and a set-point resolution of 0.1 K. Prior to each measurement, the temperature was stabilized for 2 min to ensure thermal equilibrium.

High-pressure emission spectra and decay curves were collected using a gas (nitrogen) membrane-driven diamond anvil cell (DAC; Diacell µScopeDAC-RT(G), Almax easyLab). The applied pressure was regulated using a Druck PACE 5000 controller. Ultra-low fluorescence type IIa diamonds with 0.4 mm culets were used. A stainless-steel gasket (250 µm thickness, 10 mm diameter) was pre-indented and drilled to form a 140 µm sample chamber. The sample, together with a pressure calibrant ($SrB_2O_4:Sm^{2+}$ [14]), was loaded into the drilled hole. A methanol-ethanol mixture (volume ratio 4:1) was used as the pressure-transmitting medium to ensure quasi-hydrostatic conditions during measurements.

The average lifetime of the excited state ($\tau_{avr}$) was determined by fitting the luminescence decay curves with a biexponential function according to the following equations:

$$\tau_{avr} = \frac{A_1 \tau_1^2 + A_2 \tau_2^2}{A_1 \tau_1 + A_2 \tau_2} \quad (1)$$

$$I(t) = I_0 + A_1 \cdot \exp\left(-\frac{t}{\tau_1}\right) + A_2 \cdot \exp\left(-\frac{t}{\tau_2}\right) \quad (2)$$

where $\tau_1$ and $\tau_2$ denote the individual decay components (lifetimes), and $A_1$ and $A_2$ represent the corresponding pre-exponential factors (amplitudes) of the biexponential function.

## 3. Results and discussion

The spinel-type $ZnGa_2O_4$ is a well-established host material whose crystal structure, electronic properties, and physicochemical robustness have been comprehensively documented in the literature. Owing to its outstanding thermal, chemical, and mechanical stability, $ZnGa_2O_4$ has emerged as a versatile platform for the development of functional luminescent materials. In particular, doping with transition-metal or lanthanide ions has enabled a broad range of optoelectronic and photonic applications, including persistent luminescence, photocatalysis, and lighting-related technologies.[35–40] Furthermore, the wide band gap of $ZnGa_2O_4$ (≈ 5 eV) makes it attractive for semiconductor and optoelectronic applications, underscoring its role as a multifunctional host lattice.[39] The suitability of $ZnGa_2O_4$ for advanced spectroscopic studies is further reinforced by its exceptional structural stability under extreme conditions. Notably, the $ZnGa_2O_4$ framework retains its crystallographic stability under compression up to approximately 34 GPa. Above this pressure range, a reversible pressure-induced phase transition from the cubic spinel structure to a tetragonal *I*41/*amd* phase has been reported. Upon further compression, a second structural transformation occurs at around 55 GPa, leading to the formation of a marokite-type phase with *Pbcm* symmetry.[41] This remarkable robustness under high-pressure conditions up to 30 GPa renders $ZnGa_2O_4$ particularly well suited for systematic investigations of pressure-dependent luminescence phenomena.

Structurally, $ZnGa_2O_4$ crystallizes as a normal (non-inverted) cubic spinel (space group $Fd\bar{3}m$), in which $Zn^{2+}$ ions occupy tetrahedral sites, whereas $Ga^{3+}$ ions are octahedrally coordinated (Figure 2a). These gallium-based octahedra are of central importance to the present study, as they constitute preferential substitution sites for $Ni^{2+}$ and $Cr^{3+}$ dopant ions due to their matching coordination geometry and closely comparable ionic radii ($r_{(Ga^{3+})}$ = 0.62 Å, $r_{(Cr^{3+})}$ = 0.615 Å, $r_{(Ni^{2+})}$ = 0.69 Å).[42] As evidenced by the X-ray diffraction patterns shown in Figure 2b and Figure S1, incorporation of dopant concentrations up to 1% $Ni^{2+}$ and 1% $Cr^{3+}$ does not induce any detectable structural distortions, and a single-phase spinel structure is obtained. In addition, energy-dispersive X-ray spectroscopy (EDS) elemental mapping confirms a homogeneous spatial distribution of all constituent elements within the crystallites (Figure 2c). Such compositional uniformity excludes phase segregation or dopant clustering effects, thereby ensuring the reliability, reproducibility, and intrinsic character of the evaluated luminescent properties.

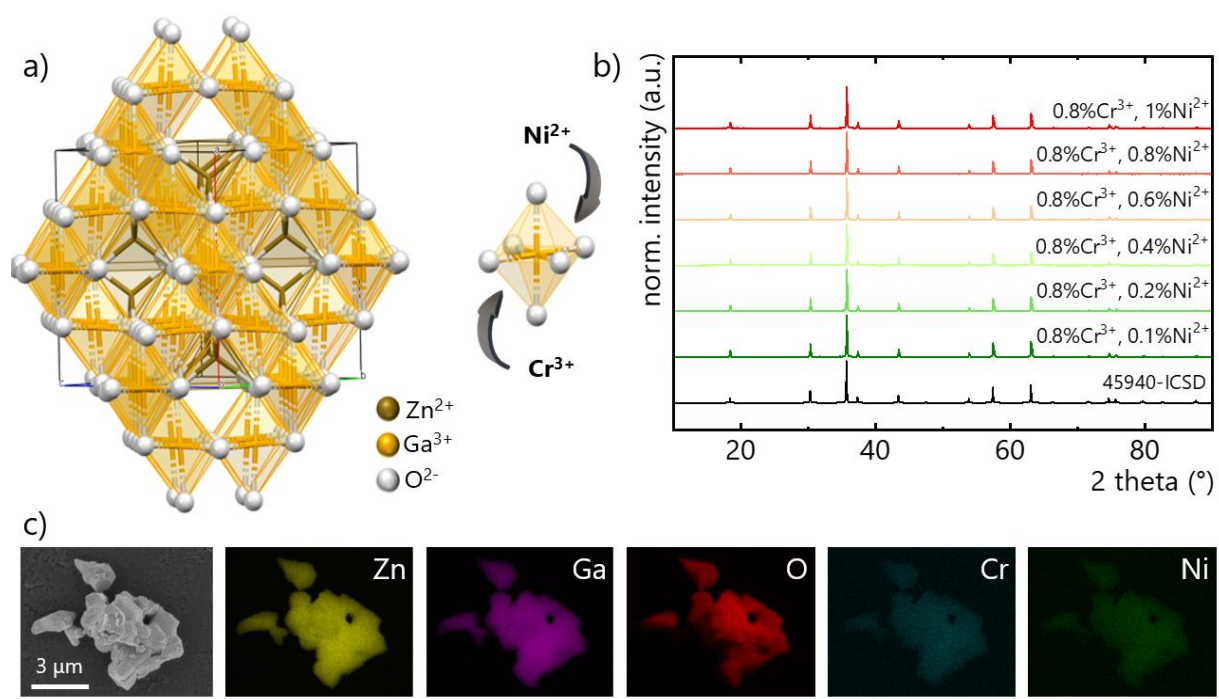

**Figure 2.** Schematic visualization of the ZnGa$_2$O$_4$ spinel structure highlighting the (GaO$_6$)$^{7-}$ octahedra, which serve as the substitution sites for Ni$^{2+}$ and Cr$^{3+}$ ions -a). X-ray diffraction patterns of ZnGa$_2$O$_4$ co-doped with Cr$^{3+}$ and Ni$^{2+}$ ions compared with the reference pattern No. 45940 from the ICSD database -b). Representative SEM micrograph and corresponding EDS elemental maps of Zn, Ga, O, Ni, and Cr for of ZnGa$_2$O$_4$:0.8% Ni$^{2+}$, 0.8% Cr$^{3+}$ -c).

Ni$^{2+}$ ions belong to the class of transition-metal ions with a $d^8$ electronic configuration, whose luminescent properties can be rationally interpreted on the basis of the Tanabe-Sugano diagram.[31,32] According to this model, the relative energies of the excited electronic states are governed by the interplay between crystal field strength and interelectronic repulsion parameters. In particular, the energy of the $^1E_g$ excited level remains nearly insensitive to variations in the crystal field strength, as it is primarily determined by electron-electron interactions rather than by metal-ligand bonding. In contrast, the $^3T_{2g}$ excited-state of $Ni^{2+}$ ions exhibits a pronounced dependence on the crystal field strength being especially sensitive to changes in the local coordination environment.[31] Variations in metal-ligand bond lengths, coordination geometry, or external perturbations such as pressure directly modulate the crystal field splitting, leading to substantial shifts of the energy of the $^3T_{2g}$ level relative to the ground $^3A_{2g}$ state. As a consequence, even subtle modifications of the host lattice can induce pronounced changes in the emission energy, bandwidth, and radiative behavior of $Ni^{2+}$-based phosphors.[34,43] This exceptional sensitivity of the $^3T_{2g}$ level to crystal field fluctuations underlies the strong responsiveness of $Ni^{2+}$ luminescence to external stimuli and constitutes the fundamental basis for exploiting $Ni^{2+}$-doped materials in pressure- and environment-sensitive optical sensing applications. As a consequence, the luminescent properties of $Ni^{2+}$-doped phosphors are strongly governed by external factors such as applied pressure as well as by the structural and chemical composition of the host matrix.[34,43–45] This pronounced crystal-field dependence results in distinct emission band shift and shape depending on the magnitude of the ligand field acting on the $Ni^{2+}$ ions. For instance, in hosts providing tetrahedral coordination for $Ni^{2+}$ ions, a sharp emission band has been reported, associated with the $^3T_2 \rightarrow {}^3T_1$ intra-configurational electronic transition.[46] In contrast, when $Ni^{2+}$ occupies an octahedral site, a broad emission band is typically observed, originating from the $^3T_{2g} \rightarrow {}^3A_{2g}$ electronic transition.[28,30,47,48]

In the case of $Cr^{3+}$ ions, the emission characteristics are also critically governed by the crystal-field strength. According to the Tanabe-Sugano diagram for octahedrally-coordinated ions with $d^3$ electronic configuration, when the *Dq/B* ratio falls below ~2.2, corresponding to a weak crystal field regime, the lowest excited state is $^4T_{2g}$.[49] Radiative relaxation from this level to the $^4A_2$ ground state gives rise to a broad emission band, characteristic of a spin-allowed transition.[50,51] Conversely, in a strong crystal field environment ($Dq/B > \sim2.2$), the $^2E$ level becomes the lowest excited state. In this regime, emission is dominated by the $^2E_g \rightarrow {}^4A_{2g}$ transition, which results in a sharp band, due to its spin-forbidden character. This dual emission behavior renders $Cr^{3+}$ ions exceptionally attractive from an application-oriented perspective, as the spectral characteristics can be rationally tuned through host lattice engineering to meet specific functional requirements.[52,53]

While the emission of $Cr^{3+}$ ions has been extensively studied and thoroughly described in the literature, particularly in the context of their practical luminescence applications, the practical implementation of $Ni^{2+}$-based phosphors remains severely limited mainly due to their intrinsically low emission intensity. Reported photoluminescence quantum efficiencies typically remain at very low levels.[29,54] To overcome this fundamental limitation, several strategies aimed at enhancing $Ni^{2+}$ emission efficiency have been proposed. One approach relies on the use of flux agents during synthesis, which can improve crystallinity; for instance, the introduction of appropriate flux additives has been shown to increase emission intensity by more than an order of magnitude. Alternatively, charge compensation strategies have been employed, whereby aliovalent ions are introduced to compensate for the charge mismatch between $Ni^{2+}$ and the substituted lattice cation. Such compensation using $Sn^{4+}$ in $LaAlO_3:Ni^{2+}$ has been demonstrated to enhance internal quantum efficiency from 0.005% up to 20.7%.[29] It was achieved by suppressing non-radiative recombination pathways associated with charge imbalance-induced defects. Another powerful strategy involves sensitization through energy transfer processes, effectively increasing the absorption cross-section of the $Ni^{2+}$ centers. Among the reported sensitizers, $Cr^{3+}$ ions have emerged as one of the most efficient candidates.[33,34,44,47,55] This superiority arises primarily from two fundamental factors: their strong absorption cross-section and favorable energy level alignment with $Ni^{2+}$ centers. In co-doped systems, optical excitation is primarily absorbed by the $Cr^{3+}$ ions or $Cr^{3+}$ and $Ni^{2+}$, followed by non-radiative energy transfer from the $^2E_g$ level of $Cr^{3+}$ to the excited states of $Ni^{2+}$ (Figure 3a). Subsequent relaxation within the $Ni^{2+}$ energy manifold leads to radiative emission in the near-infrared region. This sensitization mechanism has been experimentally shown to enhance the external quantum efficiency of $Ni^{2+}$ emission 4% to 62%.[56] Importantly, the incorporation of $Cr^{3+}$ ions typically requires only low dopant concentrations, which minimizes structural perturbations of the host lattice. This stands in contrast to flux agents or charge-compensating additives, which often need to be introduced in substantially higher amounts and may inadvertently generate additional defects or induce undesired phase changes. Consequently, $Cr^{3+}$-assisted sensitization represents a particularly attractive and structurally non-invasive route toward the realization of efficient $Ni^{2+}$-based near-infrared phosphors. When phosphors are employed as pressure sensors, rational material design is of paramount importance, as structural defects may undergo irreversible or partially irreversible rearrangements under compression. Such defect-related processes can include local lattice distortions, defect migration, or pressure-induced stabilization of metastable configurations, which may persist after decompression. As a result, the luminescence response does not necessarily return to its original state under ambient conditions, leading to hysteresis effects and compromised sensor reversibility. To minimize these adverse effects, the host lattice must exhibit high structural rigidity and resistance to defect generation under pressure.

In the case of $ZnGa_2O_4$, co-doping with $Cr^{3+}$ ions was adopted as a deliberate strategy to enhance luminescence efficiency while preserving the structural integrity of the spinel host.[33,34] Simultaneously, $Cr^{3+}$ serves as an additional luminescent center, introducing an independent spectroscopic channel that enables dual-ion optical readout. In order to determine the optimal $Cr^{3+}$ concentration, a series of samples doped exclusively with $Cr^{3+}$ ions was synthesized, and their emission intensities were evaluated under 445 nm excitation. The results revealed that the sample containing 0.8% $Cr^{3+}$ exhibited the highest emission intensity. Consequently, co-doped $ZnGa_2O_4$ samples were prepared with a fixed $Cr^{3+}$ concentration of 0.8% and varying $Ni^{2+}$ concentrations in the range of 0.1-1%. Figure 3b presents a comparison of the room-temperature emission spectrum and two excitation spectra recorded by monitoring the $Ni^{2+}$ emission at 1275 nm and the $Cr^{3+}$ emission at 694 nm. The emission spectra clearly demonstrate the coexistence of a broad near-infrared band originating from $Ni^{2+}$ ions, associated with the $^3T_{2g} \rightarrow {}^3A_{2g}$ transition, and sharp red emission lines characteristic of $Cr^{3+}$ ions, corresponding to the $^2E_g \rightarrow {}^4A_{2g}$ electronic transition. The excitation spectra, in turn, provide direct evidence for efficient energy transfer from $Cr^{3+}$ to $Ni^{2+}$ ions. This conclusion is supported by the presence of both $Cr^{3+}$- and $Ni^{2+}$-related excitation features in the excitation spectrum recorded while monitoring the $Ni^{2+}$ emission, indicating a sensitization mechanism mediated by $Cr^{3+}$ absorption followed by partial non-radiative energy transfer.[33,47]

To identify the most suitable compositions for high-pressure investigations, the luminescent properties of all co-doped samples were systematically examined. The emission spectra of $Ni^{2+}$ ions exhibited negligible dependence on $Ni^{2+}$ concentration: both the emission maximum position (1274±4 nm) and the full width at half maximum (249.5±2 nm) remained essentially unchanged (Figure 3c). This behavior can be attributed to the similar ionic radii of $Ni^{2+}$ and $Ga^{3+}$, implying that substitution of $Ga^{3+}$ by increasing amounts of $Ni^{2+}$ does not significantly alter the average $Ni^{2+}$-$O^{2-}$ bond length or the local crystal field environment.[42] It should also be emphasized that relatively low dopant concentrations were employed in this study, with the $Ni^{2+}$ content not exceeding 1 mol%, which further minimizes structural perturbations and prevents significant modification of the crystal field strength around the $Ni^{2+}$ centers. Consistently, the excitation spectra also showed no substantial changes with increasing $Ni^{2+}$ concentration (excitation spectra monitored at 694 nm are provided in the Supporting Information, Figure S2).

In addition, the luminescence kinetics were systematically investigated. Two-dimensional maps of the normalized and logarithmic luminescence decay curves are presented in Figure 3e, while the corresponding decay curves are provided in the Supporting Information - Figure S3. As can be seen, the decay curves ($\lambda_{em}$ = 1275 nm) exhibit a clearly non-exponential character, which arises from the coexistence of multiple relaxation pathways. The average lifetime of the excited $^3T_2$ level of $Ni^{2+}$, calculated using Eq. (1) and Eq. (2), shows a monotonic shortening with increasing $Ni^{2+}$ concentration. Specifically, the lifetime decreases from 1.79 ms for the sample containing 0.1% $Ni^{2+}$ to 0.47 ms for the sample with 1% $Ni^{2+}$. This behavior is attributed to concentration quenching, which becomes progressively more pronounced at higher activator concentrations. With increasing $Ni^{2+}$ content, the average distance between neighboring $Ni^{2+}$ ions decreases, facilitating non-radiative energy migration among $Ni^{2+}$ centers and increasing the probability of energy transfer to quenching sites such as defects or surface states. As a result, the probability of non-radiative relaxation is enhanced, leading to a reduction in the observed luminescence kinetics.[28,46]

Given the absence of significant differences in the spectral characteristics and excitation behavior of samples co-doped with different $Ni^{2+}$ concentrations, the composition containing 0.8% $Ni^{2+}$ and 0.8% $Cr^{3+}$ was selected for further investigations as a function of pressure and temperature. This choice was motivated by its relatively strong emission intensity compared to samples with lower $Ni^{2+}$ concentrations, while still avoiding excessive concentration quenching effects. In contrast, the sample containing 1% $Ni^{2+}$ was excluded from subsequent studies due to noticeable laser-induced heating under diode excitation. Such spontaneous heating is particularly detrimental for luminescent manometry applications, as it may lead to uncontrolled temperature rises and, consequently, distort the pressure-dependent luminescence response.

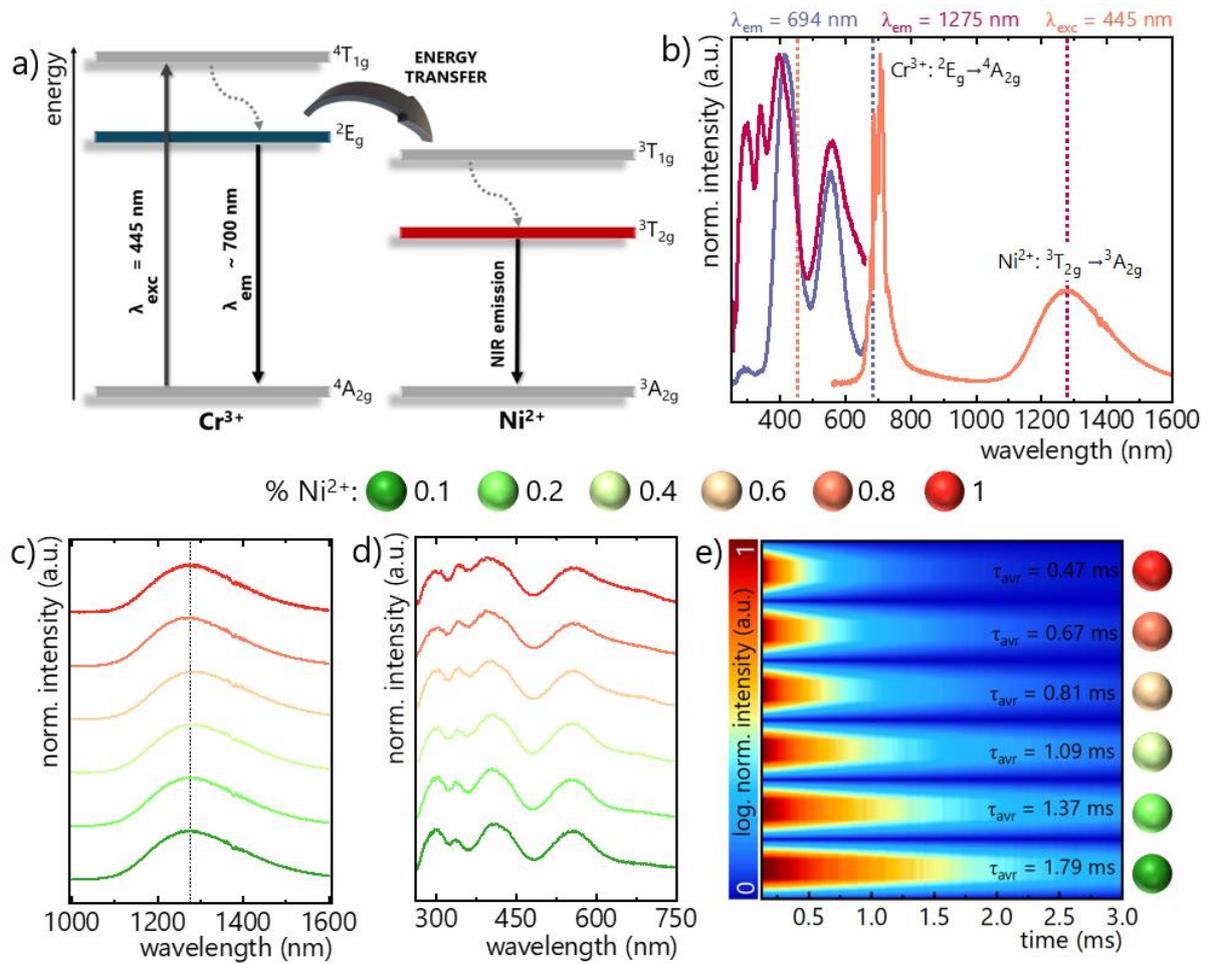

**Figure 3.** Schematic representation of the energy-level structure of a system co-doped with $Ni^{2+}$ and $Cr^{3+}$ ions, with the energy-transfer pathway indicated - a). Room-temperature emission spectrum ($\lambda_{exc}$ = 445 nm) together with excitation spectra recorded by monitoring the $Cr^{3+}$ emission ($\lambda_{em}$ = 694 nm) and $Ni^{2+}$ emission ($\lambda_{em}$ = 1275 nm) - b). Comparison of room-temperature $Ni^{2+}$ emission spectra -c), excitation spectra ($\lambda_{exc}$ = 1275 nm) -d) and maps of normalized logarithmic luminescence decay curves ($\lambda_{em}$ = 1275, $\lambda_{exc}$ = 445 nm) with calculated $\tau_{avr}$ -e) of $ZnGa_2O_4$ co-doped with 0.8% $Cr^{3+}$ and different $Ni^{2+}$ concentrations.

To evaluate the pressure-sensing capability of the $ZnGa_2O_4$:$Ni^{2+}$,$Cr^{3+}$ system, emission spectra were recorded under 445 nm excitation as a function of applied pressure up to 7.43 GPa (Figure 4a). All spectra were normalized to the $Cr^{3+}$ emission intensity to enable direct comparison of pressure-induced changes in the $Cr^{3+}$/$Ni^{2+}$ response. With increasing pressure, a pronounced reduction in the intensity of the $Ni^{2+}$ $^3T_{2g}\rightarrow{}^3A_{2g}$ broadband emission was observed relative to the sharp $Cr^{3+}$ $^2E_g \rightarrow {}^4A_{2g}$ R-line emission.[33] The initial luminescence intensity ratio (*LIR*) between the $Cr^{3+}$ and $Ni^{2+}$ emission bands increased from 0.73 under ambient conditions to 1.8 at 7.43 GPa. This behavior reflects the higher susceptibility of $Ni^{2+}$ emission to pressure-induced nonradiative deactivation. Due to its orbitally allowed character and strong electron-phonon coupling, the $^3T_{2g}$ excited state of $Ni^{2+}$ is significantly more sensitive to crystal-field perturbations and local lattice distortions than the parity-forbidden $^2E_g$ state of $Cr^{3+}$. Compression enhances crystal-field strength and may simultaneously promote defect formation or defect activation, facilitating multiphonon relaxation pathways.[57] Consequently, $Ni^{2+}$ emission undergoes more pronounced quenching, whereas the $Cr^{3+}$ R-line emission remains comparatively stable.[33] In addition to intensity modulation, distinct spectral changes were observed. The full width at half maximum (FWHM) of the $Cr^{3+}$ emission decreased from 95.6 nm to 60.6 nm with increasing pressure. This narrowing is consistent with pressure-induced enhancement of crystal-field splitting, which shifts the $^4T_2$ level to higher energies relative to the $^2E_g$ state.[58,59] Consequently, the energetic separation between these states increases, leading to reduced vibronic coupling and diminished wavefunction mixing between the $^4T_2$ and $^2E_g$ levels.[59] The decreasing overlap of their wavefunctions results in a progressive suppression of the $^4T_2$-derived broad emission component, yielding a narrower and more purely $^2E_g$-character R-line emission.[18] In the case of $Ni^{2+}$, a clear blue shift of the $^3T_{2g} \rightarrow {}^3A_{2g}$ emission band was detected (Figure 4b). The emission maximum shifted by around 83 nm over the investigated pressure range and exhibited a linear dependence on pressure with a slope of 10.7 nm GPa$^{-1}$. This behavior is consistent with the increase of the crystal field strength under compression, which raises the energy of the $^3T_{2g}$ excited state and results in a systematic spectral shift toward shorter wavelengths.[28,32,47]

The distinct and complementary pressure responses of $Cr^{3+}$ and $Ni^{2+}$ emissions enabled the development of two ratiometric pressure readout strategies (Figure 4c). In luminescence manometry, the *LIR* parameter is typically constructed either from the ratio of two narrow emission lines or from two spectral regions extracted from a broadband emission. In the present case, the spectral arrangement allows a combination of both concepts. In the first readout mode, the *LIR* is based on the ratio between a narrow spectral window centered on the $Cr^{3+}$ R-line emission and a narrow spectral window corresponding to the $Ni^{2+}$ emission band. This cross-ion approach exploits the differential pressure sensitivity of the two emitting centers and benefits from negligible spectral overlap, ensuring high discrimination and stability of the reference signal. In the second mode, the pressure is determined exclusively from the $Ni^{2+}$ emission within the NIR by calculating the ratio of intensities collected in two 5 nm-wide spectral gates positioned at pressure-sensitive regions of the $Ni^{2+}$ band. Accordingly, the *LIR* parameters were calculated according to:

$$LIR_1 = \frac{\int_{690nm}^{695nm} I(Cr^{3+}: {}^4T_{2g} \to {}^4A_{2g})d\lambda}{\int_{1325nm}^{1330nm} I(Ni^{2+}: {}^3T_{2g} \to {}^3A_{2g})d\lambda} \qquad (3)$$

$$LIR_2 = \frac{\int_{1100nm}^{1105nm} I(Ni^{2+}: {}^3T_{2g} \to {}^3A_{2g})d\lambda}{\int_{1325nm}^{1330nm} I(Ni^{2+}: {}^3T_{2g} \to {}^3A_{2g})d\lambda} \qquad (4)$$

The calibration curves derived for both readout modes (Figure 4d) reveal distinct pressure-dependent behaviors. In the mode based on $Cr^{3+}$ and $Ni^{2+}$ emissions, *LIR* increased from 1.68 at ambient pressure to 10.64 at 7.43 GPa. In contrast, the NIR-based $Ni^{2+}$ mode exhibited almost 8-fold increase of *LIR* across the same pressure range. The relative sensitivity $S_R$ was determined using equation below:

$$S_{R(p)} = \frac{1}{LIR}\frac{\Delta LIR}{\Delta p}100\% \qquad (5)$$

where $\Delta LIR$ denotes the change in the $LIR$ corresponding to a pressure variation of $\Delta p$. The maximum $S_R$ obtained for the dual-ion mode reached 27.8 % GPa$^{-1}$ at around 7.4 GPa, whereas the Ni$^{2+}$-only mode exhibited its highest sensitivity of 37.6 % GPa$^{-1}$ at ambient pressure (Figure 4e). Generally, the sensitivity of the $LIR_2$-MODE decreases monotonically with increasing pressure, while that of the $LIR_1$-MODE increases with pressure. At approximately 4 GPa, both modes exhibit comparable $S_R$ of about 26% GPa$^{-1}$. On the other hand, at 7.4 GPa, the cross-ion mode provides a $S_R$ of approximately 27.8% GPa$^{-1}$, whereas the Ni$^{2+}$-based NIR mode reaches nearly 17.2% GPa$^{-1}$.

The availability of two complementary readout strategies significantly enhances the versatility of the proposed luminescent manometer. When the phosphor is incorporated into functional matrices such as paints, coatings, or polymer composites, the NIR-based readout mode offers distinct advantages.[27] Such materials frequently exhibit strong absorption and/or autofluorescence in the visible spectral range, which may distort or attenuate visible emission signals. In contrast, operation in the near-infrared region minimizes matrix-induced optical losses and parasitic emission, enabling more reliable signal extraction in practical, application-relevant environments. Conversely, in applications requiring enhanced discrimination at elevated pressures, the dual-ion mode may provide improved robustness. Importantly, the use of narrow spectral gates as small as 5 nm suggests strong potential for two-dimensional pressure mapping using imaging systems equipped with appropriate band-pass filters, whose bandwidths typically range from several to several tens of nanometers. This feature opens a pathway toward spatially resolved pressure sensing based on simple dual-channel imaging architectures.

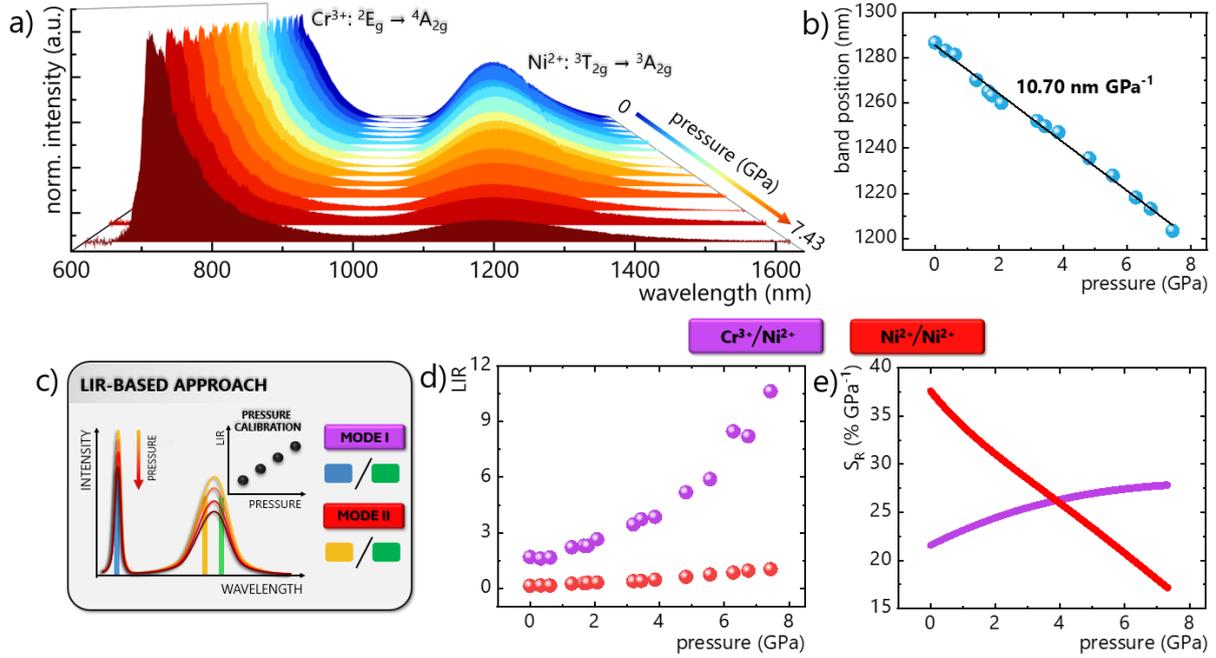

**Figure 4.** Pressure-dependent emission spectra of $ZnGa_2O_4$:0.8%$Cr^{3+}$,0.8%$Ni^{2+}$ recorded upon $\lambda_{exc}$ = 445 nm excitation in the pressure range from ambient conditions up to 7.43 GPa - a). Pressure-induced shift of the $Ni^{2+}$ $^3T_{2g} \rightarrow {}^3A_{2g}$ emission band maximum as a function of applied pressure - b). Schematic illustration of the strategy for developing the *LIR*-based luminescent manometer proposed in this work - c). Calibration curves derived from the two *LIR*-based readout modes - d) and the corresponding $S_R$ as a function of pressure - e).

Luminescence kinetics constitute a particularly powerful readout modality in luminescence manometry, primarily due to their intrinsic insensitivity to external media and optical artifacts such as scattering or absorption. This feature is of critical importance for expanding the range of potential applications, especially in heterogeneous environments or composite systems. In conventional kinetic-based luminescent pressure sensors, the pressure readout is derived from the dependence of the excited-state lifetime - most commonly the average lifetime (Figure 5a) - on the applied pressure, with this relationship serving as the calibration curve of the manometer.[12,60] While lifetime-based approach retains several important advantages, such as reduced sensitivity to fluctuations in excitation power, independence from optical alignment, and improved robustness against intensity drifts caused by scattering or absorption losses, they also exhibit inherent limitations in pressure-sensing applications.[25,26] In particular, the excited-state lifetime is frequently sensitive not only to pressure but also to temperature variations. Because radiative and non-radiative relaxation rates are both temperature dependent, even minor thermal fluctuations can significantly modify the measured decay time.[16,61] In transition-metal- and lanthanide-doped systems, increasing temperature usually leads to lifetime shortening that may mimic or obscure pressure-induced effects. As a result, lifetime-based pressure readout can suffer from cross-sensitivity, requiring either strict thermal stabilization or the implementation of additional correction protocols.[16] This intrinsic coupling between pressure and temperature responses constitutes a fundamental limitation of purely kinetic approaches in environments where thermal conditions cannot be precisely controlled. This drawback is particularly detrimental in high-pressure experiments, where pressure-induced changes in temperature occur spontaneously and are difficult to control, leading to distorted pressure readouts and reduced measurement reliability.

An alternative lifetime-based strategy, first introduced in our previous work on luminescent pressure sensors, relies on the ratio of emission intensities recorded within selected time gates of the luminescence decay (Figure 5b)[13]. This approach offers a decisive advantage: beyond providing high pressure sensitivity, it has been reported to enable pressure readout that is largely insensitive to temperature variations. Moreover, in contrast to pressure determination based on decay-time measurements - which must be performed point by point - the time-gated methodology enables two-dimensional spatial mapping of pressure distributions. Owing to the strong potential of this concept, the present study extends kinetic-based manometry by introducing a new readout strategy that simultaneously exploits time-gated emission and the pressure-dependent kinetics of two distinct luminescent ions, as schematically illustrated in Figure 5c.

Accordingly, luminescence decay curves were recorded for $Ni^{2+}$ ions by monitoring emission at 1275 nm (Figure 5d) and for $Cr^{3+}$ ions by monitoring emission at 710 nm (Figure 5e). For $Ni^{2+}$ ions, a pronounced shortening of the $^3T_2$ excited-state lifetime is observed with increasing pressure. Within the Tanabe-Sugano diagram for a $d^8$ configuration in octahedral symmetry[31,62], lattice compression increases the crystal-field strength and thus the *Dq/B* ratio, which in the present case amounts to around 1.5 under ambient conditions, regarding literature data[30]. As the crystal field strengthens, the $^3T_2$ level shifts to higher energy. According to the Tanabe-Sugano diagram, the next higher-lying state is $^1E$, which should be characterized by a longer lifetime due to its spin-forbidden character and reduced radiative transition probability. With increasing pressure, the energy separation between the $^3T_2$ and $^1E$ states decreases, promoting enhanced wavefunction overlap and partial state mixing. As a result, this could in principle enhance their interaction and increase the admixture of the longer-lived $^1E$ state. In a simplified picture, such mixing might be expected to result in the prolongation of the $^3T_2$ state lifetime. However, the experimentally observed monotonic lifetime shortening clearly indicates that this effect does not dominate the pressure response. The overall reduction of the $Ni^{2+}$ lifetime should therefore be considered as the net result of multiple competing mechanisms. In addition to state mixing within the Tanabe-Sugano framework, pressure enhances electron-phonon coupling, increases the probability of multiphonon relaxation, and may activate or amplify defect-assisted nonradiative pathways. $Ni^{2+}$ ions are particularly susceptible to such quenching processes because their $^3T_{2g} \rightarrow ^3A_{2g}$ transition is both spin-allowed and orbitally allowed, leading to strong coupling with lattice vibrations and significant vibronic broadening. In contrast to parity-forbidden transitions typical of many lanthanide ions, the excited states of $Ni^{2+}$ interact efficiently with phonons, making them inherently more sensitive to structural distortions and pressure-induced lattice stiffening.[31,41,48,62] Consequently, even moderate increases in crystal-field strength and phonon energies can substantially increase nonradiative decay rates.[63] Thus, the pressure-induced lifetime shortening reflects the cumulative effect of enhanced $^3T_{2g}$-$^1E$ mixing, increased multiphonon relaxation efficiency, and defect-assisted quenching processes. Overall, the $Ni^{2+}$ lifetime decreases approximately linearly over the investigated pressure range, from 0.59 ms at ambient pressure to 0.29 ms at 7.43 GPa, confirming the high kinetic sensitivity of $Ni^{2+}$ emission to lattice compression. Prolongation of $Ni^{2+}$ decay time as a function of pressure was earlier reported in literature.[28] In contrast, $Cr^{3+}$ ions exhibit the opposite trend: the luminescence lifetime increases from 1.00 ms to 2.09 ms with increasing pressure. This behavior can be attributed to the reduced wavefunction overlap between the $^4T_{2g}$ and $^2E_g$ states under compression, which weakens spin-orbit coupling-induced mixing between these levels. As a consequence, the contribution of the longer-lived $^2E$ state becomes more pronounced, leading to an overall prolongation in the observed decay time. Similar trends have been widely reported in the literature also for $Cr^{3+}$-doped materials subjected to compression.[61,64]

Beyond the use of the average lifetime of $^3T_{2g}$ excited state of $Ni^{2+}$ for pressure monitoring, additional readout strategies based on time-gated detection were also developed, as explained above and schematically illustrated in Figures 5b and 5c. Accordingly the integral areas of the decay curves were calculated within selected time windows: 15-20 ms for $Cr^{3+}$ ions, and 0-1 ms and 1-2 ms for $Ni^{2+}$ ions. The selection of the temporal gates was deliberately optimized to maximize the pressure-induced variation of the integrated emission signal, thereby enhancing the achievable relative sensitivity. The pressure dependence of these integrated intensities is presented in Figure 5g. For $Ni^{2+}$, the intensity decrease by factors of 2.26 and 3.6 in the respective time windows, whereas for $Cr^{3+}$ the integrated intensity increases by a factor of 5.93 up to approximately 7.4 GPa. Based on these results, two kinetic-based calibration modes were proposed, based on time-gated luminescence intensity ratios (*TG-LIRs*):

$$TG-LIR_1 = \frac{\int_{0ms}^{1ms} I(Ni^{2+}:{}^3T_{2g} \to {}^3A_{2g})dt}{\int_{1ms}^{2ms} I(Ni^{2+}:{}^3T_{2g} \to {}^3A_{2g})dt} \quad (6)$$

$$TG-LIR_2 = \frac{\int_{15ms}^{20ms} I(Cr^{3+}:{}^4T_{2g} \to {}^4A_{2g})dt}{\int_{1ms}^{2ms} I(Ni^{2+}:{}^3T_{2g} \to {}^3A_{2g})dt} \quad (7)$$

First mode relies on the ratio of integrated intensities obtained from two time gates of the $Ni^{2+}$ decay, whereas second mode is based on the ratio of integrated intensities derived from the $Ni^{2+}$ and $Cr^{3+}$ decay curves. In both cases, the calibration curves exhibit a strictly monotonic increase with pressure, with change from 3.94 to 6.52 for *TG-LIR$_1$* and from 0.04 to 0.82 for *TG-LIR$_2$*. The manometric performance of these readout strategies was quantitatively evaluated by calculating the relative sensitivity over the investigated pressure range, as below:

$$S_{R(p)} = \frac{1}{\tau_{avr}} \frac{\Delta \tau_{avr}}{\Delta p} 100\% \quad (8)$$

$$S_{R(p)} = \frac{1}{TG-LIR} \frac{\Delta TG-LIR}{\Delta p} 100\% \quad (9)$$

where *ΔTG-LIR/Δτ$_{avr}$* denote the change in the *TG-LIR/τ$_{avr}$* corresponding to a pressure variation of *Δp*. For the conventional approach based on the $\tau_{avr}$ of $Ni^{2+}$, a maximum $S_R$ of 14.17% GPa$^{-1}$ was obtained at a pressure of 7.4 GPa. By contrast, exploiting the $Ni^{2+}$ decay curve through time-gated integration resulted in the maximum sensitivity of 11.70% GPa$^{-1}$ at around 3 GPa. The highest performance, however, was achieved using the dual-ion time-gated readout, which yielded a record $S_R$ of 148.33% GPa$^{-1}$. This exceptional response originates from the opposite monotonic pressure behavior of $\tau_{avr}$ of $Cr^{3+}$ and $Ni^{2+}$: while the $Cr^{3+}$ decay time prolong upon compression, the $Ni^{2+}$ lifetime shortens. The resulting divergence of their kinetic behaviors amplifies the time-gated intensity contrast, leading to a strongly enhanced pressure-dependent response. Notably, this mode maintained a higher sensitivity across the entire investigated pressure range than the maximum sensitivities obtained for the other two readout strategies, demonstrating its superior overall performance and robustness for pressure sensing applications.

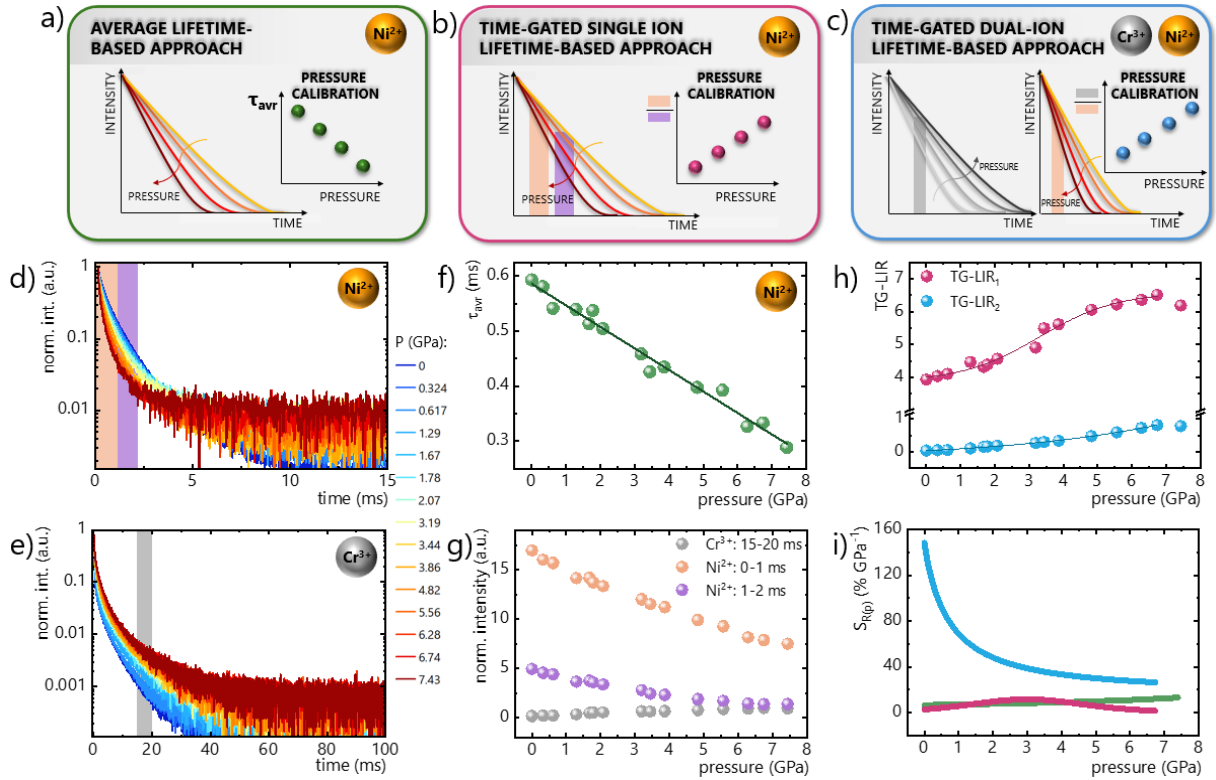

**Figure 5.** Schematic representation of the three pressure-sensing methods based on luminescence kinetics: average lifetime approach - a), time-gated single-ion lifetime approach - b), and time-gated dual-ion lifetime approach - c). Pressure-dependent decay curves of $Ni^{2+}$ ($\lambda_{em}$ = 1275 nm) - d) and $Cr^{3+}$ ($\lambda_{em}$ = 710 nm) - e) recorded upon 445 nm excitation. Pressure evolution of the $t_{avr}$ of $Ni^{2+}$ - f), and normalized emission intensities extracted from selected time gates - g). Pressure calibration curves derived from *TG-LIR* - h), and the corresponding $S_{R(p)}$ for all lifetime-based readout modes - i).

Fluctuations in pressure are frequently accompanied by spontaneous temperature changes, particularly in technologically relevant or industrial environments, where mechanical compression, friction, rapid gas expansion, or variable heat dissipation may induce unintended thermal gradients. Such coupled thermo-mechanical effects pose a significant challenge for the reliable operation of optical pressure sensors. As discussed above, increasing temperature generally enhances phonon population and strengthens electron-phonon interactions, which significantly affects the luminescence behavior of transition-metal-doped systems.[65] Elevated temperatures activate additional nonradiative pathways, leading to emission quenching, spectral redistribution, and modifications of decay dynamics. Consequently, temperature-induced changes in luminescence intensity, spectral shape, or lifetime may partially mimic or obscure pressure-induced effects, introducing cross-sensitivity into luminescence-based pressure readout schemes. In practical applications, this may necessitate strict thermal stabilization or the implementation of compensation protocols. Therefore, a thorough evaluation of the temperature response of the developed pressure sensor is essential in order to assess its real-world applicability beyond controlled laboratory conditions. To this end, temperature-dependent emission spectra (Figure 6a) as well as luminescence decay curves for $Ni^{2+}$ (Figure 6b) and $Cr^{3+}$ (Figure S5) were recorded. Upon increasing temperature, a progressive decrease in emission intensity was observed for both ions. However, the quenching of $Ni^{2+}$ emission was markedly stronger, which is consistent with its orbitally allowed $^3T_{2g} \rightarrow {}^3A_{2g}$ transition and strong electron-phonon coupling.[62,66] In contrast, the parity-forbidden $^2E_g \rightarrow {}^4A_{2g}$ emission of $Cr^{3+}$ exhibits inherently weaker coupling to lattice vibrations and thus greater thermal robustness.[33] This differential thermal response directly affects ratiometric readout schemes. To quantify the temperature cross-sensitivity, calibration curves were constructed using the same *LIR* as those applied for pressure sensing (Equations (3) and (4)). The corresponding calibration curves are presented in Figure 6b. For the $Cr^{3+}/Ni^{2+}$ *LIR* mode, the *LIR* value changed from 2.18 to 50.9 between 93 K and 473 K. Interestingly, the temperature evolution was non-monotonic: an initial decrease in *LIR* was observed with increasing temperature, followed by a plateau around 193 K, and subsequently an increase at higher temperatures. This behavior reflects the interplay between differential thermal quenching rates and possible thermal population redistribution within the excited-state manifolds. In contrast, when only the $Ni^{2+}$ emission band was utilized, an approximately 12-fold increase in *LIR* was observed over the temperature range from 93 K to 473 K, indicating high thermometric sensitivity. The relative temperature sensitivity $S_{R(T)}$ was calculated analogously to the pressure case using:

$$S_{R(T)} = \frac{1}{LIR}\frac{\Delta LIR}{\Delta T}100\% \qquad (10)$$

where *ΔLIR* denotes the change in the *LIR* corresponding to a temperature variation of *ΔT*. The values obtained (Figure 6c) indicate that at room temperature the $S_R$ reaches approximately 2.09% K$^{-1}$ for the $Cr^{3+}/Ni^{2+}$ mode and 1.63% K$^{-1}$ for the $Ni^{2+}$-based mode. When monitoring the ratio between the $Cr^{3+}$ and $Ni^{2+}$ emission bands ensures $S_R$ exceeding 1% K$^{-1}$ over a broad high-temperature range from 280 K to 473 K. In contrast, the ratiometric analysis based solely on $Ni^{2+}$ emission band provides sensitivity above 1% K$^{-1}$ in the lower temperature region from 93 K to 300 K. Relative sensitivities above 1% K$^{-1}$ are generally regarded as a benchmark for reliable luminescent thermometry. Accordingly, the investigated $ZnGa_2O_4:Cr^{3+},Ni^{2+}$ system demonstrates clear potential as an efficient optical thermometer. Additionally, the coexistence of the two various temperature ranges with $S_R$ above 1% K$^{-1}$ within a single material platform effectively extends the thermometric working range across nearly the entire 93-473 K interval. This multimodal response demonstrates that the system combines high sensitivity with exceptional operational versatility, establishing it as a broadly applicable luminescent thermometer.

Moreover, the temperature dependence of the pressure readout was systematically evaluated by determining the Thermal Invariability Manometric Factor (*TIMF*), defined as the ratio of pressure sensitivity to temperature sensitivity:

$$TIMF = \frac{S_{R(p)}}{S_{R(T)}} \qquad (11)$$

For both readout modes, *TIMF* remains relatively low across the investigated pressure range, indicating a pronounced thermometric response of the material (Figure 6d). This behavior is consistent with the high temperature sensitivity observed in the corresponding thermometric calibration curves. Nevertheless, the $Ni^{2+}$-only mode exhibits around fourfold higher *TIMF* value at ambient pressure, and at 7.43 GPa it is approximately 1.5-fold higher than for the $Cr^{3+}$-$Ni^{2+}$ mode. This indicates a greater susceptibility of the $Ni^{2+}$-based pressure readout to temperature-induced interference. At the same time, this enhanced temperature responsiveness creates an opportunity to deliberately exploit the *LIR*-based modes for the design of a highly sensitive luminescent thermometer.

An analogous analysis was performed for kinetic readout modes. The average lifetime of the $Ni^{2+}$ $^3T_{2g}$ excited state decreases from 1.36 ms to 0.25 ms over the temperature range from 93 K to 433 K (Figure 6f), confirming strong thermal sensitivity of the $Ni^{2+}$ decay dynamics. Time-gated emission intensities were extracted using the same temporal windows as defined for pressure measurements (Figure S6), and temperature-dependent *TG-LIR* calibration curves were proposed (Figure 6g). For the $Ni^{2+}$-based time-gated mode, *TG-LIR* increases from 1.62 to 11.51 between 93 K and 433 K, indicating pronounced thermometric response. In contrast, for the $Cr^{3+}/Ni^{2+}$ time-gated dual-ion mode, *TG-LIR* decreased by approximately 3-times between 93 K and 313 K, demonstrating significantly reduced temperature sensitivity. Relative temperature sensitivities were again determined for all three kinetic-based readout modes (Figure 6h). For both the $Ni^{2+}$ *TG-LIR* mode and the average lifetime approach, the maximum $S_{R(T)}$ values were observed at temperatures around 347 K and 433 K, reaching approximately 1.33% $K^{-1}$ and 1.12% $K^{-1}$, respectively. In contrast, the $Cr^{3+}/Ni^{2+}$ dual-ion time-gated mode exhibits the lowest maximum temperature sensitivity, approximately 1.01 % $K^{-1}$, occurring at lower temperature around 165 K. When combined with the previously determined pressure sensitivities, this analysis reveals that the dual-ion time-gated mode achieves the highest *TIMF* value of 1059.5 K GPa$^{-1}$, confirming its superior resistance to temperature-induced artifacts (Figure 6i). In other words, this readout strategy enables reliable pressure monitoring with minimal cross-sensitivity to thermal fluctuations. Conversely, the Ni$^{2+}$-single ion modes, both *t$_{avr}$*- and *TG-LIR*-based, display lower *TIMF* values (ranging from 11.22 to 22.15 K GPa$^{-1}$ and 11.49 to 1.59 K GPa$^{-1}$, respectively, across the investigated pressure range), reflecting their strong intrinsic thermometric character.

Overall, the combined thermometric and manometric characterization of the ZnGa$_2$O$_4$:Cr$^{3+}$, Ni$^{2+}$ system reveals an important multifunctional aspect. Beyond serving as a highly sensitive luminescent manometer, the material also offers the possibility of dual-parameter sensing. By appropriate selection of spectral or temporal gates, one can preferentially enhance either pressure or temperature sensitivity. In particular, the Ni$^{2+}$-based time-gated and proposed *LIR*-based approaches provide high temperature sensitivity with moderate pressure response, while the dual-ion Cr$^{3+}$/Ni$^{2+}$ time-gated strategy enables pressure monitoring that is largely immune to temperature fluctuations. To date, only a limited number of phosphor systems have been reported that provide two genuinely independent readout channels for simultaneous pressure and temperature sensing.[5,22,67–71] The demonstrated decoupling of pressure and temperature sensitivities markedly broadens the application potential of the present system in scenarios where both parameters vary simultaneously, including industrial process monitoring, high-load mechanical systems, pressurized reactors and pipelines, and other harsh or dynamically changing environments in which reliable pressure readout must remain robust against unintended thermal fluctuations.

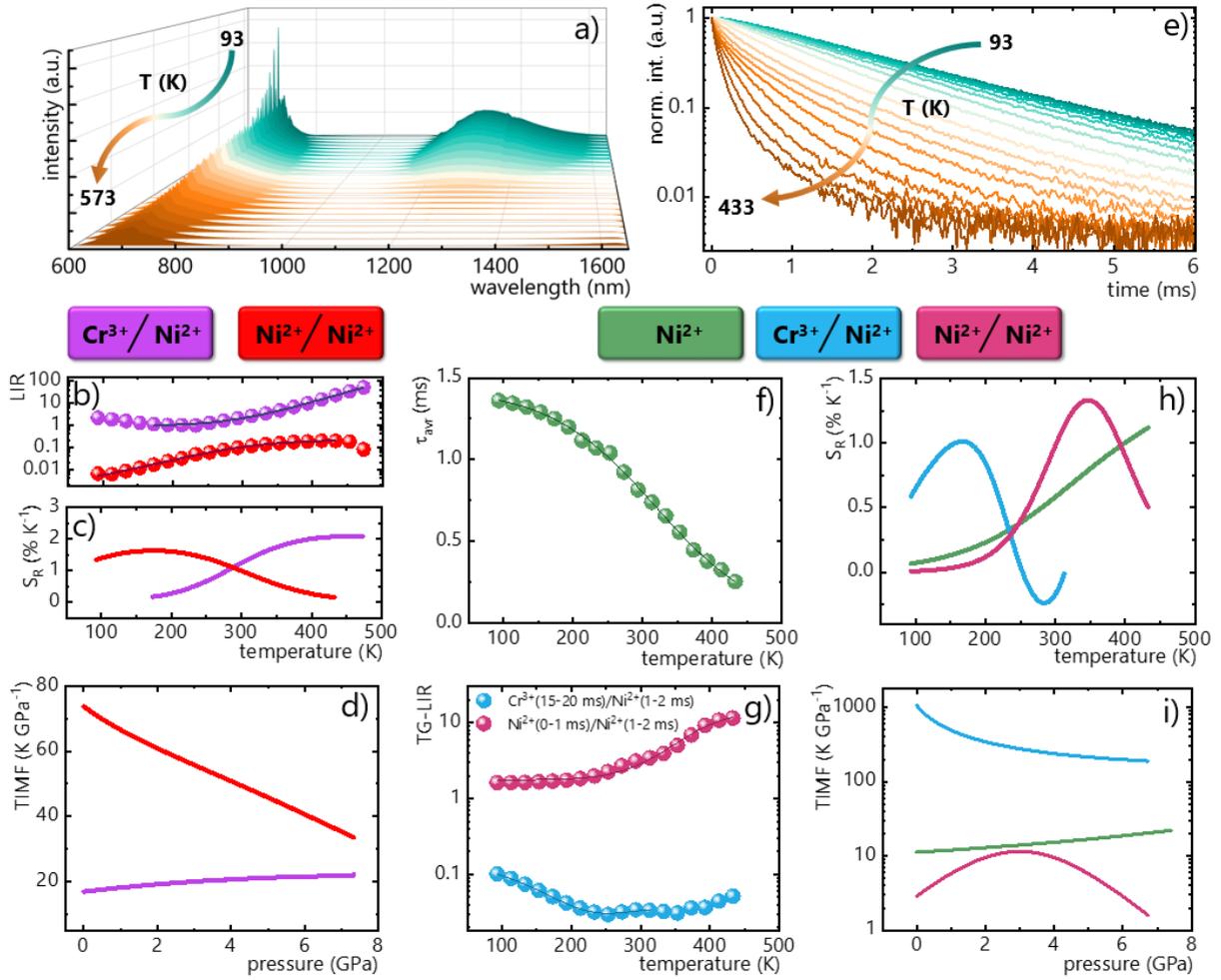

**Figure 6.** Temperature-dependent emission spectra recorded under $\lambda_{exc}$ = 445 nm - a). Thermal evolution of two calibration curves based on *LIR* - b), together with the corresponding relative sensitivity $S_R$ - c) and *TIMF* as a function of pressure - d). Luminescence decay curves ($\lambda_{exc}$ = 445 nm, $\lambda_{em}$ = 1275 nm) measured as a function of temperature - e), and the calculated $\tau_{avr}$ of $Ni^{2+}$ - f). Calibration curves based on the *TG-LIR* method, defined as the intensity ratio of $Cr^{3+}$ (15-20 ms gate) to $Ni^{2+}$ (1-2 ms gate), as well as the intensity ratio of $Ni^{2+}$ emission (0-1 ms to 1-2 ms gates) - g), with the corresponding $S_R$ - h) and *TIMF* as a function of pressure - i).

While luminescence thermometry has evolved into a mature and extensively benchmarked field[72], luminescence manometry remains comparatively sparse and less systematically evaluated. The limited number of reported systems and the diversity of readout strategies necessitate careful benchmarking to position new concepts within the broader landscape.[73] Against this background, the manometric performance of $ZnGa_2O_4:Cr^{3+},Ni^{2+}$ was critically
29

assessed relative to state-of-the-art luminescent pressure sensors. Figure 7a summarizes ratiometric systems operating on *LIR* calibration curves based on emission spectra, including only those exhibiting relative sensitivities exceeding 1% GPa$^{-1}$.[4–6,8–13,17–19,68,74] Within this highly selective group, $ZnGa_2O_4:Cr^{3+},Ni^{2+}$ ranks among the most sensitive *LIR*-based luminescent manometers reported so far. Notably, the comparison highlights a fundamental gap in the field: only a limited number of pressure sensors operate in the near-infrared spectral region (>800 nm).[9,11,12,17] This spectral window is highly advantageous for practical applications, offering reduced optical scattering, enhanced penetration depth in optically dense media, and compatibility with NIR-optimized detection systems.[75] In this context, the $ZnGa_2O_4:Cr^{3+},Ni^{2+}$ platform extends luminescent manometry toward technologically relevant NIR operation. On the other hand, Figure 7b further compares lifetime-based manometric systems with reported *TIMF* values.[4,6,12,13,17,60,76] Strikingly, the dual-ion time-gated decay ratio approach, introduced here for the first time, enables record-high relative pressure sensitivity. Importantly, this exceptional sensitivity is accompanied by a high value of the *TIMF* value, confirming that pressure readout remains effectively decoupled from temperature variations. Thus, the proposed kinetic strategy combines ultrahigh sensitivity with intrinsic thermal robustness, overcoming a long-standing limitation of lifetime-based pressure sensing. It should be noted, however, that while dual-ion *TG-LIR*-based approach delivers exceptionally high pressure sensitivity, its practical implementation requires a more advanced detection architecture. In particular, simultaneous acquisition of signals from distinct time gates associated with pressure- and temperature-sensitive emission channels typically necessitates multiple synchronized detection pathways or high-speed time-resolved instrumentation. This increases both system complexity and overall cost, potentially limiting large-scale or industrial deployment. Nevertheless, several technological solutions have been proposed in the literature to mitigate these limitations. Time-multiplexed detection schemes, in which sequential temporal windows are rapidly sampled using a single high-speed detector combined with programmable gating electronics, enable accurate separation of the relevant luminescence components without the need for parallel detection channels.[77–83] These advances open the possibility for cost-effective implementation of time-gated dual-ion sensing platforms while preserving their intrinsic advantages, including high sensitivity, temperature decoupling, and robustness against intensity fluctuations.

Beyond single-parameter performance benchmarking, consideration of simultaneous pressure–temperature sensing reveals an even more distinctive advantage of the proposed system. As highlighted in recent review article[84], only a limited number of systems have been reported to date that enable simultaneous optical monitoring of both variables[4–8,15,67,69–71,85]. In this landscape, the $ZnGa_2O_4$:$Cr^{3+}$,$Ni^{2+}$ platform advances the field by introducing a fundamentally distinct operational concept based on dual, independently addressable calibration pathways. For the first time, two independent calibration curves based on the time-gated *LIR* approach are implemented for decoupled pressure and temperature readout. This dual-calibration architecture enables simultaneous monitoring of both thermodynamic parameters while preserving high sensitivity in each channel and maintaining minimal cross-interference. Beyond the *TG-LIR* strategy, the system additionally supports a conventional *LIR*-based approach for temperature sensing, providing an alternative spectral readout pathway. The coexistence of kinetic and *LIR*-based ratiometric modes within a single material platform establishes an exceptional level of operational flexibility. Such multimodal adaptability renders $ZnGa_2O_4$:$Cr^{3+}$,$Ni^{2+}$ an unusually versatile sensing architecture, capable of tailoring the readout strategy to the demands of specific experimental or application environments.

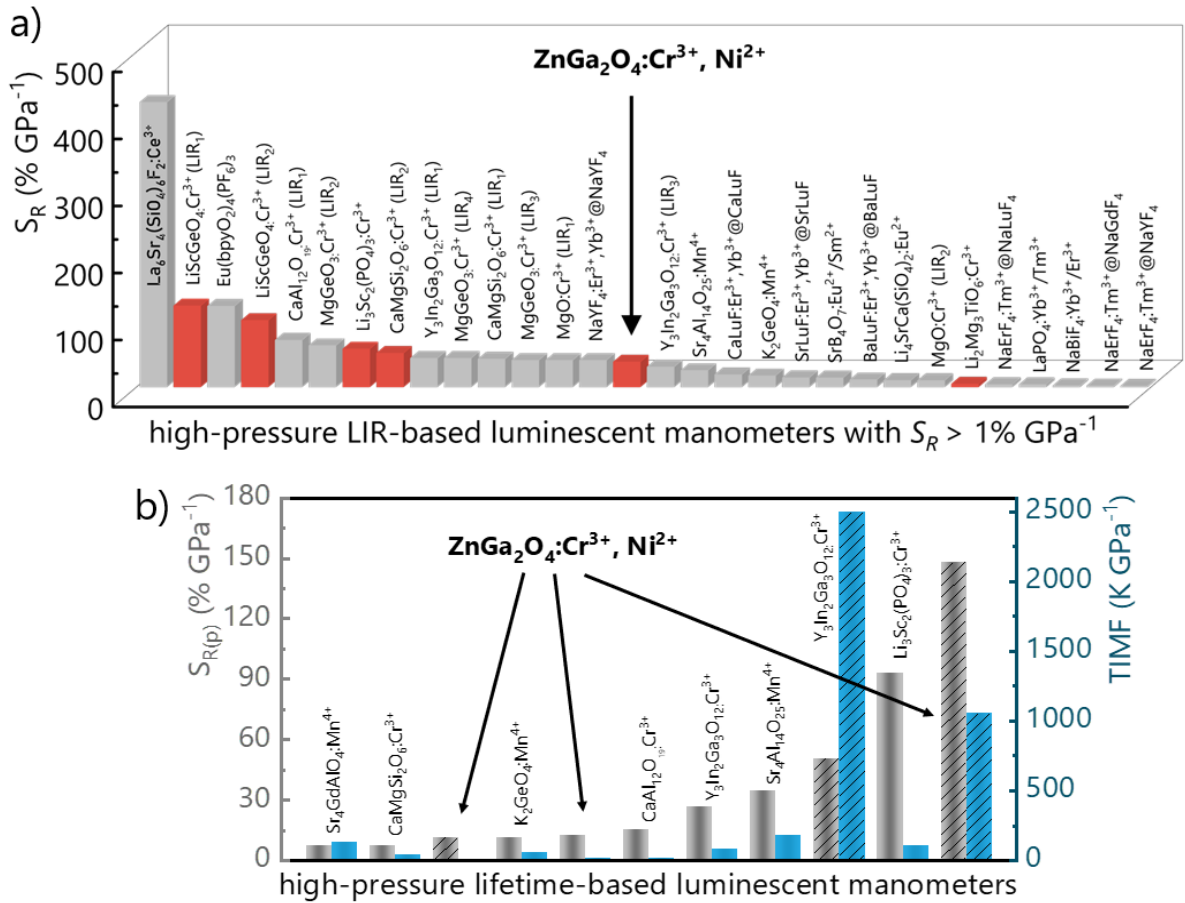

**Figure 7.** Comparison of the manometric performance of reported luminescent pressure sensors operating in: ratiometric *LIR*-based mode with relative sensitivity exceeding 1% GPa$^{-1}$ (red bars correspond to sensors operating in the near-infrared spectral region >800 nm) -a) and lifetime-based mode (dashed-filled bars indicate sensors employing the time-gated approach -b).

**Conclusions**

This work demonstrates a synergistic interaction between Ni$^{2+}$ and Cr$^{3+}$ ions as a new class of active centers for advanced luminescent pressure-temperature sensing in the ZnGa$_2$O$_4$:Ni$^{2+}$,Cr$^{3+}$ spinel lattice. Application of hydrostatic pressure up to around 7.4 GPa induces an exceptionally large blueshift of the Ni$^{2+}$ emission band associated with the $^3T_{2g} \rightarrow {}^3A_{2g}$ transition. In contrast, the Cr$^{3+}$ $^2E_g \rightarrow {}^4A_{2g}$ emission remains relatively spectrally stable, thereby providing an intrinsic reference channel. The combination of a highly pressure-sensitive Ni$^{2+}$ emission and a comparatively stable $Cr^{3+}$ emission forms the physical basis for the multimodal sensing concept proposed here. Exploiting this pronounced differential response, a multifunctional optical manometer was developed. Two principal pressure-readout strategies were implemented: (i) spectral ratiometric modes based on the luminescence intensity ratio constructed either from selected regions of the $Ni^{2+}$ emission band or from the $Cr^{3+}/Ni^{2+}$ emissions and (ii) time-gated ratiometric modes utilizing pressure-induced redistribution of emission intensity within defined temporal windows of $Ni^{2+}$ and $Cr^{3+}$-$Ni^{2+}$ emission. Among the investigated strategies, the dual-ion time-gated $Cr^{3+}$-$Ni^{2+}$ mode delivered the highest overall performance, achieving a record relative pressure sensitivity of $S_{R(p)}$ = 148.33% $GPa^{-1}$ while simultaneously maintaining negligible susceptibility to temperature fluctuations, as evidenced by a high *TIMF* value of 1059.5 K $GPa^{-1}$. This effective decoupling of thermo- and manometric responses represents a critical advance over conventional lifetime-based pressure sensors, where cross-sensitivity to temperature typically limits practical applicability. Notably, $ZnGa_2O_4$:$Ni^{2+}$,$Cr^{3+}$ currently represents the most sensitive luminescent manometer operating in lifetime-based mode reported to date, combining record-high pressure sensitivity with operation in the NIR spectral window. Moreover, this work demonstrates for the first time the utilization of $Ni^{2+}$ ions as active centers for high-performance optical pressure sensing, as well as the deliberate exploitation of $Ni^{2+}$-$Cr^{3+}$ synergistic interaction to enhance and stabilize the pressure readout. The introduction of $Ni^{2+}$ and its cooperative behavior with $Cr^{3+}$ thus establishes a new design paradigm for transition-metal-based luminescent manometers. Importantly, the same material platform was further engineered to function as a combined thermometer-manometer. By appropriate selection of spectral bands and temporal detection windows, the relative sensitivities toward pressure and temperature can be selectively enhanced. In this context, temperature monitoring was found to be most effective using the time-gated $Ni^{2+}$ emission mode, which achieved a maximum relative temperature sensitivity of $S_{R(T)}$ = 1.33% $K^{-1}$ at 347 K, accompanied by a *TIMF* value of 11.49

K GPa$^{-1}$, confirming its strong thermometric character. Additionally, a pronounced thermometric response was demonstrated for *LIR*-based modes, including both Ni$^{2+}$-only and Cr$^{3+}$-Ni$^{2+}$ approaches. Monitoring the intensity ratio between the Cr$^{3+}$ and Ni$^{2+}$ emission bands ensures a relative sensitivity exceeding 1% K$^{-1}$ over a broad high-temperature range (280-473 K). In contrast, the Ni$^{2+}$-based ratiometric mode provides sensitivity above 1% K$^{-1}$ in the lower temperature region (93-300 K). Thus, the ZnGa$_2$O$_4$:Ni$^{2+}$,Cr$^{3+}$ system enables universal, multi-parameter sensing within a single phosphor, allowing either quasi-independent readout channels or simultaneous monitoring of both pressure and temperature.

Taken together, these results establish ZnGa$_2$O$_4$:Ni$^{2+}$,Cr$^{3+}$ as a highly versatile and application-oriented platform for precise luminescent pressure and temperature sensing. The availability of complementary multiple readout modes enables flexible adaptation to diverse operational requirements, ranging from cost-efficient single-parameter sensing to advanced dual-parameter monitoring in technologically demanding industrial environments.

**Acknowledgements**

This work was supported by National Science Center Poland (NCN) under project No. 2023/49/N/ST5/01020. Maja Szymczak gratefully acknowledges the support of the Foundation for Polish Science through the START program. Authors would like to acknowledge dr Damian Szymanski for SEM and EDS analyses.

**Conflict of Interest**

The authors declare no competing financial or commercial interests.